\documentclass{pasj00}

\begin{document}
\SetRunningHead{Author(s) in page-head}{Running Head}
\Received{2006/05/11}
\Accepted{2006/06/30}

\title{An Optically Dark GRB Observed by HETE-2: GRB\,051022}

\author{%
  Yujin E. \textsc{Nakagawa}\altaffilmark{1}
  Atsumasa \textsc{Yoshida}\altaffilmark{1,2}
  Satoshi \textsc{Sugita}\altaffilmark{1}
  Kaoru \textsc{Tanaka}\altaffilmark{1}
  Nobuyuki \textsc{Ishikawa}\altaffilmark{1}\\
  Toru \textsc{Tamagawa}\altaffilmark{2}
  Motoko \textsc{Suzuki}\altaffilmark{2}
  Yuji \textsc{Shirasaki}\altaffilmark{2,4}
  Nobuyuki \textsc{Kawai}\altaffilmark{2,3}
  Masaru \textsc{Matsuoka}\altaffilmark{5}\\
  Jean-Luc \textsc{Atteia}\altaffilmark{6}
  Alexandre \textsc{Pelangeon}\altaffilmark{6}
  Roland \textsc{Vanderspek}\altaffilmark{7}
  Geoff B. \textsc{Crew}\altaffilmark{7}
  Joel S. \textsc{Villasenor}\altaffilmark{7}\\
  Nat \textsc{Butler}\altaffilmark{7,8}
  John \textsc{Doty}\altaffilmark{7,9}
  George R. \textsc{Ricker}\altaffilmark{7}
  Graziella \textsc{Pizzichini}\altaffilmark{10}
  Timothy Q. \textsc{Donaghy}\altaffilmark{11}\\
  Donald Q. \textsc{Lamb}\altaffilmark{11}
  Carlo \textsc{Graziani}\altaffilmark{11}
  Rie \textsc{Sato}\altaffilmark{3}
  Miki \textsc{Maetou}\altaffilmark{1}
  Makoto \textsc{Arimoto}\altaffilmark{3}
  Jun'ichi \textsc{Kotoku}\altaffilmark{3}\\
  J. Garret \textsc{Jernigan}\altaffilmark{8}
  Takanori \textsc{Sakamoto}\altaffilmark{12}
  Jean-Francois \textsc{Olive}\altaffilmark{13}
  Michel \textsc{Boer}\altaffilmark{13}\\
  Edward E. \textsc{Fenimore}\altaffilmark{14}
  Mark \textsc{Galassi}\altaffilmark{14}
  Stanford E. \textsc{Woosley}\altaffilmark{15}
  Makoto \textsc{Yamauchi}\altaffilmark{16}\\
  Kunio \textsc{Takagishi}\altaffilmark{16}
  and
  Isamu \textsc{Hatsukade}\altaffilmark{16}
}
\altaffiltext{1}{Department of Physics and Mathematics, Aoyama Gakuin University, \\5-10-1 Fuchinobe, Sagamihara, Kanagawa 229-8558}
\email{yujin@phys.aoyama.ac.jp}
\altaffiltext{2}{The Institute of Physical and Chemical Research, 2-1 Hirosawa, Wako, Saitama 351-0198}
\altaffiltext{3}{Department of Physics, Tokyo Institute of Technology, \\2-12-1 Ookayama, Meguro-ku, Tokyo, 152-8551}
\altaffiltext{4}{National Astronomical Observatory, 2-21-1 Osawa, Mitaka, Tokyo, 181-8588}
\altaffiltext{5}{Tsukuba Space Center, JAXA, 2-1-1 Sengen, Tsukuba, Ibaraki, 305-8505}
\altaffiltext{6}{LAT, Observatoire Midi-Pyr\'{e}n\'{e}es, 14 Avenue E. Belin, F$-$31400 - Toulouse, France}
\altaffiltext{7}{Center for Space Research, MIT, 70 Vassar Street, Cambridge, Massachusetts, 02139, USA}
\altaffiltext{8}{Space Sciences Laboratory, University of California at Berkeley, \\Berkeley, California, 94720-7450, USA}
\altaffiltext{9}{Noqsi Aerospace, LTd., 2822 South Nova Road, Pine, Colorado, 80470, USA}
\altaffiltext{10}{INAF/IASF Bologna, Via Gobetti 101, 40129 Bologna, Italy}
\altaffiltext{11}{Department of Astronomy and Astrophysics, University of Chicago, \\5640 South Ellis Avenue, Chicago, Illinois 60637, USA}
\altaffiltext{12}{Goddard Space Flight Center, NASA, Greenbelt, Maryland, 20771, USA}
\altaffiltext{13}{Centre d'Etude Spatiale des Rayonnements, Observatoire Midi-Pyr\'{e}n\'{e}es, \\9 Avenue de Colonel Roche, 31028 Toulouse cedex 4, Rrance}
\altaffiltext{14}{Los Alamos National Laboratory, P. O. ox 1663, Los Alamos, NM, 87545, USA}
\altaffiltext{15}{Department of Astronomy and Astrophysics, University of California at Santa Cruz, \\477 Clark Kerr Hall, Santa Cruz, California, 95064, USA}
\altaffiltext{16}{Faculty of Engineering, Miyazaki University, Gakuen Kibanadai Nishi, Miyazaki, 889-2192}



\maketitle

\begin{abstract}
 GRB\,051022 was detected at 13:07:58 on 22 October 2005 by HETE-2.
 The location of GRB\,051022 was determined immediately by the flight 
 localization system.
 This burst contains multiple pulses and has a rather
 long duration of about 190 seconds. The detections of candidate X-ray and
 radio afterglows were reported, whereas no optical afterglow was
 found. The optical spectroscopic observations of the host galaxy
 revealed the redshift ${\rm z} = 0.8$.
 Using the data derived by HETE-2 observation of the prompt emission,
 we found the absorption 
 $N_{\rm H} = (8.8_{-2.9}^{+3.1}) \times 10^{22}$ cm$^{-2}$
 and the visual extinction $A_{V} = 49_{-16}^{+17}$ mag
 in the host galaxy.
 If this is the case, no detection of any optical transient would be
 quite reasonable.
 The absorption derived by the Swift XRT observations of the afterglow
 is fully consistent with those obtained from the early HETE-2
 observation of the prompt emission.
 Our analysis implies an interpretation that the absorbing medium
may be outside the external shock at $R \sim 10^{16}$ cm, which could be
 a dusty molecular cloud.
\end{abstract}

\section{Introduction}
Among gamma-ray bursts (GRBs), ``optically dark'' bursts are 
known as GRBs without accompanying optical transients. 
From lack of a precise location being usually determined 
by its optical afterglow, it is generally difficult to 
search out its corresponding host galaxy.  Hence there are only 
a few hosts found to date by radio counterparts for this kind 
of bursts,
and detailed studies such as their morphology or taxonomy 
are very premature.
Why are these GRBs ``optically dark''?  
The answer is still unclear.  
Their optical counterparts might have decayed much rapidly than others,
and/or could be hidden behind heavily absorbing matters.
Some works suggest that those are associated with 
dusty molecular clouds along the line of sight \citep{rei02} or 
distant GRBs with $z\gtrsim5$ \citep{fru99a, lam00}.

In soft X-ray band, measuring the absorptions in spectra of 
prompt emissions and/or afterglows from GRBs
could provide us important information about the environments around
sources.
Actually some authors reported time-variable absorptions in spectra of
some GRB prompt emissions and afterglows.
One of the interesting properties of GRB\,970828 is a time-variable absorption
in the spectra of the afterglow, which could be due to
circum-burst medium \citep{yos01}.
Similar absorption in the afterglows were also reported by \citet{owe98} and
\citet{str04}.

For prompt emissions, time-variable absorptions were reported
for GRB\,980329 \citep{fro00}, GRB\,990705 \citep{ama00} and
GRB\,010222 \citep{int01}.
Optical counterparts were found for these bursts, 
in contrast to the case for GRB\,970828.
\citet{fro00} suggested that a time-variable absorption in the spectra
of GRB\,980329 was due to
the internal shock accompanied by the expanding fireball.
For GRB\,990705, an absorption feature in the prompt emission
might be explained by a photoelectric absorption by a medium
at z=0.86 \citep{ama00}.

We report here constant absorptions throughout the
prompt emission and the afterglow of GRB\,051022 observed respectively by
HETE-2 and Swift.
We also present the localization and the spectral properties of
GRB\,051022.
We discuss output energies of the burst and 
an evidence for an intervening dense medium along the line of sight.

\section{Observations and Analyses}

\subsection{Localization}
The gamma-ray burst GRB\,051022 was detected with the three 
scientific instruments aboard HETE-2, 
the Wide-Field X-ray Monitor (WXM; 2-25 keV; \cite{shi03}), 
Soft X-ray Camera (SXC; 0.5-10 keV; \cite{vil03}) and 
French Gamma Telescope (FREGATE: 6-400 keV; \cite{att03}),  
at 13:07:58 on 22 October 2005 \citep{gra05, tan05}.
The location of GRB\,051022 was determined immediately by the flight
WXM and SXC localization system. The GCN Notices reporting the position
were sent out 45 seconds after the onset based on the WXM flight
localization, and 119 seconds after the onset based on the SXC
flight localization.

The WXM location was a circle centered at
${\rm R.A.} = \timeform{23h55m55s.2}$,
${\rm decl.} = \timeform{19D39'36''.0}$
(J2000) 
with a radius of $\timeform{5'}$ (90 \% confidence region)
by the ground analysis of the data.
The brightness of GRB\,051022 was sufficient in soft X-rays to
determine the position with an error radius down to $\timeform{1'19''.8}$
using the SXC data independently of the WXM localization.
Unfortunately the SXC data was partially lost due to the dropout of
internet connection to the ground station at Cayenne.
The SXC localization was made by the ground analysis and resulted in a
circle centered at
${\rm R.A.} = \timeform{23h56m03s.7}$,
${\rm decl.} = \timeform{19D37'10''.9}$
(J2000) 
($l = \timeform{105D27'20''.0}$,
$b = \timeform{-41D21'54''.8}$)
with a radius of $\timeform{2'30''}$ (90 \% confidence region).

The Inter-Planetary Network (IPN) also reported
its position \citep{hur05} constrained on an
annulus centered at
${\rm R.A.} = \timeform{03h11m39s}$,
${\rm decl.} = \timeform{16D32'32''}$
(J2000) 
with a radius of 46.4533$\pm$0.0684 degrees
(the quoted error is 3$\sigma$ confidence region).
The SXC error circle was just encompassed in this annulus position.

The Swift XRT instrument began observing 3.5 hours after the trigger
the field of GRB\,051022 based on the HETE-2 localization, 
and detected a bright unknown fading X-ray source \citep{rac05a},
just $\timeform{1'6''.8}$ away from 
the center of the SXC error circle.
The XRT located it with an accuracy of $\timeform{4''}$ 
and the Chandra narrowed later its error region 
down to  $\timeform{0.7''}$ \citep{pat05}, 
where the VLA observation at 8.5\,GHz discovered a bright 
radio source \citep{cam05}.
The probable host galaxy was reported by several groups 
\citep{cas05, ber05, coo05, nys05, blo05, uga05}, 
and the optical spectroscopic observation using
the 200-inch Hale Telescope
at Palomar Observatory detected a strong line at 6736 ${\rm \AA}$ which
corresponds to O$\emissiontype{II}$ 3727 ${\rm \AA}$ and determined a
redshift ${\rm z} = 0.8$ \citep{gal05}.

Using the data of Swift XRT instrument, spectral analyses
of X-ray afterglow were performed (\cite{but05a}, \yearcite{but05b}; \cite{rac05b}).
\citet{but05a} reported the absorption $N_{\rm H} = (0.84\pm0.07)
\times 10^{22}$ cm$^{-2}$ which is greater than the Galactic value
in the direction to the burst.
The light curve of X-ray afterglow presented a break
at t$_{\rm break} = 2.9\pm0.2$ days, and the jet opening
angles were estimated \citep{rac05b} if this break was due to
the sideway expansion; $\theta_{\rm jet} = 4.3$ degrees for the
HETE-2 spectral parameters \citep{dot05} and $\theta_{\rm jet} = 4.4$
degrees for the Konus-Wind spectral parameters \citep{gol05}.

\subsection{Temporal Properties}
Unfortunately, because of the dropout of internet connection
to our ground station at Cayenne at the trigger time, we lost
the time tagged photon data of the FREGATE. The spectral and
temporal analyses
of the FREGATE are performed using the 5.24 s resolution data
from this reason.
The upper five panels in Figure \ref{lc_par} show the time history
of GRB\,051022 in five energy bands, where ${\rm t} = 0$ shows the trigger time
which corresponds to 13:07:58 on 22 October 2005 UT.
The event consists of multiple pulses and has a rather long
duration $T_{\rm{90}} = \rm{178} \pm \rm{8}$ s in the 2$-$25 keV
energy band and $\rm{157} \pm \rm{5}$ s in the 30$-$400
keV energy band.

\subsection{Absorption in Spectra}
In our analysis, we use the following models: power-law (PL),
power-law times exponential cutoff
(PLE) and Band function (GRBM; \cite{ban93}).
The WXM and the FREGATE were pointed to around
${\rm R.A.} = \timeform{01h08m00s}$,
${\rm decl.} = \timeform{11D08'00''}$
(J2000) 
($l = \timeform{129D28'14''.0}$,
$b = \timeform{-51D31'40''.9}$) 
when the GRB was observed. 
The FREGATE has larger field-of-view of 70 degrees than that of 
the WXM, and we found many mildly bright soft sources at that time 
within the FREGATE field-of-view but outside the WXM 
consulting the ASM/RXTE database.
The spectrum of FREGATE shows somewhat larger counts 
near its lower energy end than those from the WXM in the same 
band. This could be due to contamination from the soft sources mentioned
above.
To avoid this discrepancy only in the lower end of energy band of 
the FREGATE, we employ data above 40 keV up to 400 keV 
for joint fits and find very good agreement in 
continuum spectra with the WXM and the FREGATE. 

\begin{figure}
  \begin{center}
    \FigureFile(80mm,114mm){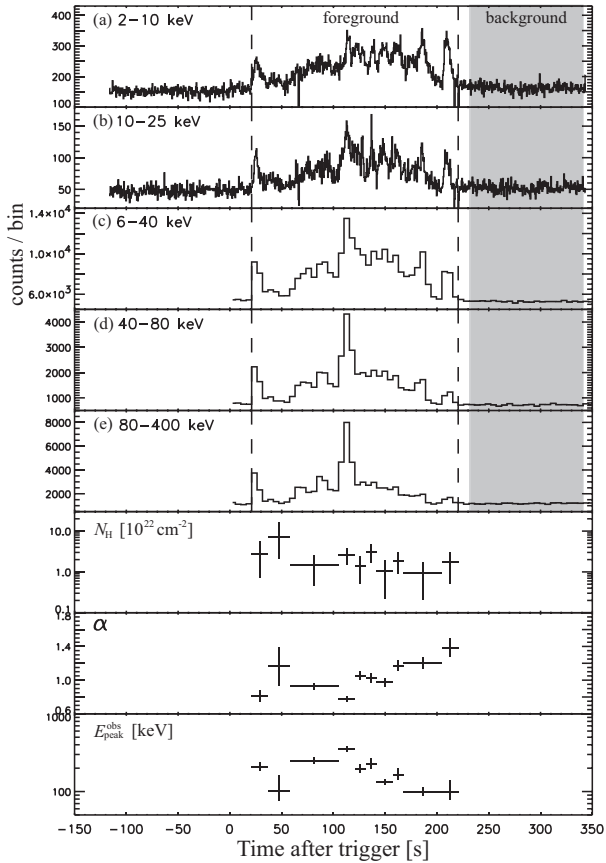}
  \end{center}
  \caption{Time history (counts/bin) of GRB\,051022 observed by the
    WXM in the 2-10 keV energy band
 (a) and the 10-25 keV energy band (b) with 0.5 s time bins; and by the FREGATE
 in the 6-40 keV energy band (c), the 40-80 keV energy band (d) and the 80-400 keV
 energy band (e) with 5.24 s time bins. The dashed lines show the foreground region
 and the hatched area displays the background region.
 The bottom three panels show the variation of spectral parameters.
 The quoted errors correspond to the 90 \% confidence region.}\label{lc_par}
\end{figure}

\begin{figure}
  \begin{center}
    \FigureFile(80mm,80mm){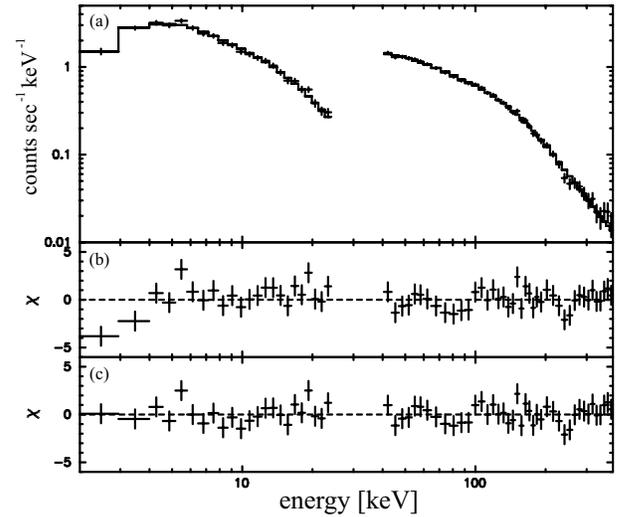}
  \end{center}
  \caption{The average spectra with the best-fit absorbed GRBM model (a),
  the residual using the unabsorbed GRBM model (b) and the residual
  using the absorbed GRBM model (c).}\label{spectra}
\end{figure}

First of all, we analyze the average spectrum of the prompt emission
using the total duration t $=$ 21$-$220 s.
For background we employ data during t $=$ 231$-$341 s.
Their time regions are indicated in Figure \ref{lc_par}.
Any unabsorbed model does not provide an acceptable fit to the data.
We find a deficit of photons in the spectrum using
unabsorbed GRBM model in lower energy band below 4 keV (see panel (b) of  
Figure \ref{spectra}).
The Galactic value of absorption in the direction of the
burst is $4.09 \times 10^{20}$ cm$^{-2}$ \citep{dic90},
which is negligible for this fit.
Then, we adopt an absorption as a free parameter and fit the data.
The fit is clearly improved (see panel (c) of Figure \ref{spectra}) and
the most favorable model is the absorbed GRBM with
$N_{\rm H}$ $=$ $(1.51_{-0.50}^{+0.53}) \times 10^{22}$ cm$^{-2}$.
Using the absorbed GRBM model, we find $\alpha = 1.01_{-0.03}^{+0.02}$,
$\beta = 1.95_{-0.14}^{+0.25}$,
$E_{\rm peak}^{\rm obs} = 213\pm18$ keV,
and fluences $S_{\rm X}$ $=$ $(21.4\pm0.2)
\times 10^{-6}$ $\rm{ergs}$ $\rm{cm^{-2}}$ (2$-$30 keV) and
$S_{\gamma}$ $=$ $(131\pm1) \times 10^{-6}$ $\rm{ergs}$
$\rm{cm^{-2}}$ (30$-$400 keV).
The quoted errors correspond to the 90 \% confidence region for
$N_{\rm{H}}$, $\alpha$, $\beta$ and $E_{\rm peak}^{\rm obs}$, and the 68
\% confidence region for $S_{\rm X}$ and $S_{\gamma}$.
Thus the ratio of fluences is log($S_{\rm X}$/$S_{\gamma}$) $=$ $-$0.786, and
GRB\,051022 is classified into the ``Classical'' hard GRB
in the HETE-2 sample \citep{sak05}.

In previous studies for GRBs, the absorption appeared 
only in a part of a
prompt emission and/or afterglow \citep{ama00, fro00, int01, yos01, str04}.
Then we perform the time resolved spectral analyses for 10
time intervals to investigate a time variation of the absorption,
and summarize the results in Table \ref{spec_tb}.
The bottom three panels in Figure \ref{lc_par} show the
time variation of these spectral parameters.
In all time intervals, we adopt the absorbed PLE model
because reliable $\beta$ cannot be
obtained from these fittings with GRBM model.
The quoted errors correspond to the 90 \% confidence region for $N_{\rm{H}}$,
$\alpha$ and $E_{\rm peak}^{\rm obs}$,
and the 68 \% confidence region for
$S_{\rm X}$ and $S_{\gamma}$.

From the spectral variation,
the initial pulse (t $=$ 21.0$-$36.7 s) is hard and
accompanied by a soft pulse (t $=$ 36.7$-$57.7 s).
During the later long phase (t $=$ 57.7$-$220 s), the spectrum shows
a softening trend.
These are consistent with the general view of the time variation of 
GRB spectra.
The most remarkable result is that $N_{\rm H}$ is significantly
needed and seems constant in all the time intervals
contrary to the results from the previous studies.
Then we perform spectral fitting by assuming the absorption
fixed to the value obtained by the analysis of the average spectrum 
and get acceptable fits for all intervals
(see $\chi'^2$ in Table \ref{spec_tb}).
In conclusion, $N_{\rm H}$ does not seem to have a
significant evolution.

\section{Discussion and Conclusion}
Using the measured redshift $z = 0.8$ \citep{gal05}, the luminosity
distance is given by $d_{\rm L} = 1.49\times10^{28}$ cm
($\Omega_m=0.32$, $\Omega_\Lambda=0.68$ and $H_{\rm 0}=72$ km s$^{-1}$ Mpc$^{-1}$).
$E_{\rm iso}$ turns out to be
$(6.6\pm1.3) \times 10^{53}$ ergs
by integrating the best-fit time integrated spectrum
in the observer frame over the energy ranging from
$1/(1+z)$ keV to $10/(1+z)$ MeV (\cite{blo01}, \yearcite{blo03}; \cite{ama02}).
We also find the peak energy of $\nu F_{\nu}$ spectrum in the source frame
$E_{\rm peak}^{\rm src} = 382_{-32}^{+33}$ keV.
These values are lying on $E_{\rm peak}^{\rm src}-E_{\rm iso}$
relation \citep{ama02}.

In the standard view, the fireball is collimated \citep{wax98, fru99b}
and the collimation corrected energies are concentrated around
$10^{51}$ ergs \citep{blo03, ghi04}.
This scenario is believed to appear as the achromatic break 
in observed afterglow light curve \citep{rho97, sar99}.
The break time of the light curve
$t_{\rm break}=2.9\pm0.2$ days was reported using the
X-ray afterglow observations \citep{rac05b}.
If this is due to the jet break,
the jet opening angle $\theta_{\rm jet}$ \citep{sar99} turns out
to be $\sim4.2$ deg assuming $n\sim0.1$ cm$^{-3}$
and $\eta_{\gamma}\sim0.2$,
where $n$ is a proton number density around the GRB site and $\eta_{\gamma}$
is an energy conversion efficiency.
Then, we estimate the collimation corrected energy $E_{\gamma}$ \citep{blo03, ghi04}
scaling $E_{\rm iso}$ by (1$-\cos\theta_{\rm jet}$)
and find $E_{\gamma}=(1.8\pm0.3) \times 10^{51}$ ergs.
This value is lying on $E_{\rm peak}^{\rm src}-E_{\gamma}$ relation
\citep{ghi04}.
Thus the jet break at 2.9 days seems plausible.

The follow-up observations of X-ray afterglow were performed
by the Swift XRT instrument from 16:35:54 on 22 October 2005 \citep{rac05a}.
Using the three orbits of XRT data, we performed spectral analyses
of the afterglow, and found
$N_{\rm H} = (0.91_{-0.11}^{+0.12}) \times 10^{22}$ cm$^{-2}$
and $\Gamma=2.0\pm0.1$ with $\chi^2/{\rm d.o.f.} = 166/150$.
These values are consistent with that previously reported by \citet{but05a}.
We also found a constant absorption during the XRT observation.
The absorption derived by the Swift XRT observation is
fully consistent with those obtained from the early HETE-2 observation
of the prompt emission.
In earlier studies, it was reported that
the absorption varied in time \citep{ama00, fro00,
int01, yos01, str04} and therefore it was interpreted as an absorption
by the circum-burst dusty medium in the very vicinity
of internal shocks.
In contrast, our analyses present the constant absorption
throughout the prompt emission and the afterglow.
It might be due to the intervening medium along the line of sight
which is in the outside of external shocks;
i.e. a molecular cloud in the host galaxy.

The large absorption would cause the extinctions of
the afterglow emission.
If the intervening medium is located somewhere in the
middle between the source and our galaxy,
the absorption should be
at least 
$N_{\rm H}$ $\gtrsim$ $(1.51_{-0.50}^{+0.53}) \times 10^{22}$ cm$^{-2}$.
We find the visual extinction at minimum $A_{V} = 8.4_{-2.8}^{+3.0}$ mag using the
relationship $A_{V} = N_{\rm H}/(1.79 \times 10^{21}$ cm$^{-2})$
found by \citet{car89}.
Then we estimate the extinctions for other wavelengths at minimum
using the extinction curves found by \citet{pre95};
$A_{U} = 13.2_{-4.4}^{+4.7}$ mag,
$A_{B} = 11.3_{-3.7}^{+4.0}$ mag, $A_{R} = 6.3_{-2.1}^{+2.2}$ mag,
$A_{I} = 4.0_{-1.3}^{+1.4}$ mag, $A_{J} = 2.4\pm0.8$ mag,
$A_{H} = 1.6_{-0.5}^{+0.6}$ mag, $A_{K} = 1.0\pm0.3$ mag and
$A_{L} = 0.5\pm0.2$ mag.
These large extinctions could explain the fact that 
no optical afterglow was found despite the prompt and deep
search \citep{tor05, cen05}.

The most extreme case is that the intervening medium is located in the
very vicinity of the host galaxy of GRB\,051022.
The absorption should be scaled by $(1+z)^{3}$
\citep{gun65}.
We find $N_{\rm H} = (8.8_{-2.9}^{+3.1}) \times 10^{22}$ cm$^{-2}$
and the extinctions are
$A_{V} = 49_{-16}^{+17}$ mag, $A_{U} = 77_{-25}^{+27}$ mag,
$A_{B} = 66_{-22}^{+23}$ mag, $A_{R} = 37_{-12}^{+13}$ mag,
$A_{I} = 24\pm8$ mag, $A_{J} = 14\pm5$ mag,
$A_{H} = 9\pm3$ mag, $A_{K} = 6\pm2$ mag and
$A_{L} = 3\pm1$ mag at maximum.
If this is the case, no detection of any optical transient
would be quite reasonable.

Considering that no optical afterglow is found despite a prompt deep
search of afterglow down to $R\sim20.0$ mag \citep{cen05},
GRB\,051022 is very similar to GRB\,970828
which is one of the important GRBs for which
no optical afterglow was found despite a prompt deep search down
to $R\sim24.5$ mag \citep{ode97}.
\citet{djo01} reports that the radio afterglow 
of GRB\,970828 is located between two bright sources
(A and B in Figure 3 of \cite{djo01}).
They suggest that there might be 
a dust lane intersecting a single galaxy or it might be a
merging system of three components (A, B and C in Figure 3
of \cite{djo01}).
The star formation rates (SFR) are reported by the authors as  
SFR$\sim$1.2 $\MO$ yr$^{-1}$ for component A and
SFR$\sim$0.3 $\MO$ yr$^{-1}$ for B.

The absorption for GRB\,970828 using the brightness of X-ray and
radio afterglow was $N_{\rm H}\gtrsim6 \times 10^{21}$ cm$^{-2}$
in the source frame, therefore a dusty molecular cloud is one
possible interpretation \citep{djo01}.
Meantime \citet{yos01} shows a time-variable absorption with
$N_{\rm H} = 3.13 \times 10^{22}$ cm$^{-2}$ in the source frame
based on the spectral analyses of the X-ray afterglow.
Therefore the most probable interpretation is that the absorption
could be dominantly due to the medium near the GRB site \citep{yos01, djo01}.

For GRB\,051022, several authors report, based on the optical and 
IR observations, the most probable candidate of 
its host galaxy \citep{cas05, ber05, coo05, nys05, blo05, gal05, uga05}
at the location consistent with those of the burst (SXC), 
the X-ray afterglow (XRT), and the radio transient (VLA).
From these images, 
the host galaxy (i.e., galaxy ``B'' in the above references) 
seems roundly extended at least $\sim \timeform{1''}$ in radius, 
which corresponds to be about 7\,kpc at $z=0.8$, 
greater than the typical size of a galactic bulge.  
Therefore it would not be an ``edge-on'' spiral galaxy.

There is also a report that this galaxy is blue with 
SFR of more than 20 $\MO$ yr$^{-1}$ \citep{cas06}.
The value is far larger than that for GRB\,970828.
This large SFR could be consistent with a dusty molecular cloud
in the galaxy.
In addition, GRB\,051022 shows the constant absorption 
in the soft X-ray band 
throughout the prompt emission and the afterglow.
This sharply contrasts with the previous results 
\citep{ama00, fro00, int01, yos01, str04}.
The absorption for GRB\,051022 is evaluated to be
$N_{\rm H} = (8.8_{-2.9}^{+3.1}) \times 10^{22}$ cm$^{-2}$,
which is larger than that of GRB\,970828.
Our results favor an interpretation that the absorbing medium
is outside the external shock at $R \gtrsim 10^{16}$ cm
and could be a dusty molecular cloud.

\bigskip
We would like to thank the HETE-2 members for their support.
We acknowledge the use of public data from the Swift data archive.
The HETE-2 mission is supported in the US by NASA contract NASW-4690; in
Japan in part by the Ministry of Education, Culture, Sports, Science,
and Technology Grant-in-Aid 14079102; and in France by CNES contract
793-01-8479.
One of the authors (Y.E.N.) is supported by the JSPS Research
Fellowships for Young Scientists.

%

\begin{longtable}{llllrrll}
 \caption{Spectral model parameters for the time resolved spectra of GRB\,051022.}\label{spec_tb}
 \hline\hline
 Time Region & $N_{\rm{H}}$\footnotemark[$*$] & $\alpha$ & 
 $E_{\rm peak}^{\rm obs}$\footnotemark[$\dagger$] &
 \multicolumn{1}{l}{$S_{\rm X}$\footnotemark[$\ddagger$]} &
 \multicolumn{1}{l}{$S_{\gamma}$\footnotemark[$\ddagger$]} &
 $\chi^2$ (d.o.f.) & $\chi'^2$\footnotemark[$\S$] (d.o.f.) \\
 (s) &  &  & (keV) &  &  &  & \\
 \hline
 \endfirsthead
 \hline\hline
 Time Region & $N_{\rm{H}}$\footnotemark[$*$] & $\alpha$ & 
 $E_{\rm peak}^{\rm obs}$\footnotemark[$\dagger$] &
 \multicolumn{1}{l}{$S_{\rm X}$\footnotemark[$\ddagger$]} &
 \multicolumn{1}{l}{$S_{\gamma}$\footnotemark[$\ddagger$]} &
 $\chi^2$ (d.o.f.) & $\chi'^2$\footnotemark[$\S$] (d.o.f.) \\
 (s) &  &  & (keV) &  &  &  & \\
 \hline
 \endhead
 \hline
 \multicolumn{8}{l}{\hbox to 0pt{\parbox{180mm}{\footnotesize
     \footnotemark[$*$] $N_{\rm H}$ denotes the photoelectric
     absorption in units of $10^{22}$ cm$^{-2}$ with 90 \% confidence level errors.
     \par\noindent
     \footnotemark[$\dagger$] $E_{\rm peak}^{\rm obs}$ denotes the peak energy of
     $\nu F_{\nu}$ spectrum in observer frame with 90 \% confidence level errors.
     \par\noindent
     \footnotemark[$\ddagger$] $S_{\rm X}$ and $S_{\gamma}$ denote the fluences
     in the energy range 2$-$30 keV and 30$-$400 keV respectively, in units of $10^{-6}$\\ ergs
     $\rm{cm^{-2}}$ with 68 \% confidence level errors.
     \par\noindent
     \footnotemark[$\S$] $\chi'^2$ denotes the chi-square of the fits, assuming
     the absorption fixed to the value obtained by the analysis of\\ the average spectra.
 }}}
 \endfoot
 \hline
 \multicolumn{8}{l}{\hbox to 0pt{\parbox{180mm}{\footnotesize
     \footnotemark[$*$] $N_{\rm H}$ denotes the photoelectric
     absorption in units of $10^{22}$ cm$^{-2}$ with 90 \% confidence level errors.
     \par\noindent
     \footnotemark[$\dagger$] $E_{\rm peak}^{\rm obs}$ denotes the peak energy of
     $\nu F_{\nu}$ spectrum in observer frame with 90 \% confidence level errors.
     \par\noindent
     \footnotemark[$\ddagger$] $S_{\rm X}$ and $S_{\gamma}$ denote the fluences
     in the energy range 2$-$30 keV and 30$-$400 keV respectively, in units of $10^{-6}$\\ ergs
     $\rm{cm^{-2}}$ with 68 \% confidence level errors.
     \par\noindent
     \footnotemark[$\S$] $\chi'^2$ denotes the chi-square of the fits, assuming
     the absorption fixed to the value obtained by the analysis\\ of the average spectra.
 }}}
 \endlastfoot
21.0-36.7 & $2.7_{-2.0}^{+2.7}$ & $0.81_{-0.07}^{+0.07}$ & $210_{-23}^{+30}$ & $1.32_{-0.04}^{+0.06}$ & $10.8_{-0.3}^{+0.2}$ & 79.1 (106) & 80.0 (107) \\
36.7-57.7 & $7.1_{-5.0}^{+8.8}$ & $1.16_{-0.23}^{+0.23}$ & $101_{-25}^{+60}$ & $0.61_{-0.05}^{+0.03}$ & $1.8_{-0.4}^{+0.1}$ & 52.6 (57) & 56.2 (58) \\
57.7-105 & $1.5_{-1.0}^{+1.1}$ & $0.93_{-0.04}^{+0.04}$ & $248_{-21}^{+26}$ & $4.34_{-0.07}^{+0.09}$ & $32.8_{-0.6}^{+0.6}$ & 82.2 (106) & 82.2 (107) \\
105-121 & $2.6_{-1.1}^{+1.3}$ & $0.77_{-0.03}^{+0.03}$ & $352_{-25}^{+29}$ & $3.27_{-0.06}^{+0.05}$ & $41.1_{-0.5}^{+0.4}$ & 99.1 (106) & 102 (107) \\
121-131 & $1.4_{-0.9}^{+1.0}$ & $1.05_{-0.05}^{+0.05}$ & $195_{-22}^{+29}$ & $1.68_{-0.04}^{+0.04}$ & $8.8_{-0.3}^{+0.2}$ & 83.4 (106) & 83.4 (107) \\
131-142 & $3.0_{-1.4}^{+1.7}$ & $1.02_{-0.06}^{+0.06}$ & $226_{-30}^{+40}$ & $1.60_{-0.04}^{+0.03}$ & $9.9_{-0.3}^{+0.2}$ & 87.2 (106) & 90.4 (107) \\
142-157 & $1.1_{-0.8}^{+0.9}$ & $0.98_{-0.05}^{+0.05}$ & $132_{-10}^{+12}$ & $2.37_{-0.04}^{+0.05}$ & $9.9_{-0.4}^{+0.2}$ & 65.9 (106) & 66.6 (107) \\
157-168 & $1.9_{-1.0}^{+1.1}$ & $1.17_{-0.06}^{+0.06}$ & $165_{-24}^{+34}$ & $1.58_{-0.03}^{+0.04}$ & $6.2_{-0.3}^{+0.2}$ & 61.4 (88) & 61.8 (89) \\
168-205 & $1.0_{-0.7}^{+0.8}$ & $1.21_{-0.07}^{+0.06}$ & $100_{-11}^{+15}$ & $3.23_{-0.07}^{+0.05}$ & $8.1_{-0.3}^{+0.2}$ & 78.4 (106) & 79.7 (107) \\
205-220 & $1.7_{-1.2}^{+1.4}$ & $1.39_{-0.11}^{+0.11}$ & $100_{-21}^{+42}$ & $1.18_{-0.05}^{+0.03}$ & $2.5_{-0.3}^{+0.1}$ & 82.1 (90) & 82.2 (91) \\
\end{longtable}

\end{document}